\documentclass[10pt,twoside,BCOR7mm,DIV15,headinclude,footexclude,
               cleardoubleempty,idxtotoc]
{scrartcl}

\usepackage{natbib}
\usepackage[font=small,labelfont=bf]{caption}
\usepackage[english]{babel}
\usepackage{graphicx}
\usepackage{hyperref}
\usepackage{scrpage2}
\usepackage{ifthen}
\usepackage{booktabs}
\usepackage{amsmath}
\usepackage{amssymb}
\usepackage{multicol}
\usepackage{float}
\usepackage{hyperref}

\hypersetup{breaklinks=true
,colorlinks=true,linkcolor=black,urlcolor=blue
,citecolor=black}

\addto\captionsenglish{%
}

\makeatletter
\renewcommand{\@biblabel}[1]{}
\renewcommand{\@cite}[2]{%
{#1\ifthenelse{\boolean{@tempswa}}{,#2}{}}}
\makeatother
\ofoot{\thepage}
\ifoot{}

\setcapindent{0em}
\setheadsepline{1pt}

\setkomafont{pagehead}{\normalfont\sffamily}
\setkomafont{pagenumber}{\normalfont\rmfamily}

\makeatletter
\newcommand{\listofcontributions}{\@starttoc{con}}

\newcommand{\l@contribution} {\@dottedtocline{1}{1.5em}{2.3em}}
\makeatother

\newenvironment{contribution}{
\setcounter{section}{0}
\setcounter{figure}{0}
\setcounter{table}{0}
}{
\newpage
\lehead{}
\rohead{}
}

\begin{document}

\setlength{\baselineskip}{2.5ex}

\begin{contribution}

\newcommand{\fdg}{\mbox{\ensuremath{.\!\!^\circ}}}

\lehead{P.\ Massey, K.\ F.\ Neugent, \& N.\ Morrell}

\rohead{Finding WRs in the Local Group: How and Why}

\begin{center}
{\LARGE \bf Finding Wolf-Rayet Stars in the Local Group}
\medskip

{\it\bf P. Massey$^1$, K. F. Neugent$^1$, \& N. Morrell$^2$}\\

{\it $^1$Lowell Observatory, Flagstaff, AZ, USA}\\
{\it $^2$Las Campanas Observatory, La Serena, Chile}

\begin{abstract}
We summarize past and current surveys for WRs among the Local Group galaxies, emphasizing both the why and how.  Such studies are invaluable for helping us learn about massive star evolution, and for providing sensitive tests of the stellar evolution models.  But for such surveys to be useful, the completeness limits must be well understood.  We illustrate that point by following the ``evolution" of the observed WC/WN ratio in nearby galaxies. We end by examining our new survey for WR stars in the Magellanic Clouds, which has revealed a new type of WN star, never before seen. 
 
\end{abstract}
\end{center}

\begin{multicols}{2}

\section{Introduction}

The motivation for WR surveys should be to improve our understanding of massive star evolution, and not just for the sake of adding a few more WRs to our lists.  Stellar evolutionary modeling is
hard, and there are numerous simplifications that our theoretician colleagues have had to adopt.   How good are these approximations?   Until we compare observations with the model predictions, we don't know. 
But, to succeed at this our surveys must be complete enough to be useful.  

The basic premise of the ``Conti scenario" \citep{Conti75} is that a massive star strips off its H-rich outer layers through stellar winds, first revealing He and N, the products of the CNO cycle.  If there is sufficient additional mass loss, then the star peels down far enough for us to see C and O, the productions of triple-$\alpha$ He burning.  Today we would argue that this process is occasionally aided by Roche-lobe overflow in close binaries and/or episodic mass loss during the LBV stage \citep{SmithOwocki06}, although whether these processes are important for most massive stars or not remains an open question. 

These stellar winds are driven by radiation pressure in highly ionized metal lines, and hence the mass-loss rates are metallicity dependent.  Since the nearby star-forming galaxies cover a range of $20\times$ in metallicity \citep{MasseyARAA}, these galaxies make ideal laboratories for studying massive star evolution as a function of metallicity. 

We show in Fig.~\ref{fig:rsg} an example of what we can learn from such studies.  We have plotted the $\log$ of the relative number of RSGs and WRs against the metal abundance, measured from the oxygen content of H\,{\sc ii} regions, for four Local Group galaxies for which we believe the numbers are relatively complete.  Note that this quantity changes by an order of magnitude over a similar change in metallicity.  \citet{Maeder80} was the first to suggest that this number ratio should be very sensitive to the (initial) metallicity of the stars, and that a comparison with observations would be a good test of the models.

If our surveys for WRs and other evolved stars are to be useful in learning about massive star evolution, we must be careful that they are complete, or at least that their limitations are well understood.   To be useful, WR surveys must be sensitive enough to detect the weakest-lined WRs at the faintest magnitude limits one expects to find them, and be volume-limited, and not magnitude-limited.  
\begin{figure}[H]
\begin{center}
\includegraphics[width=0.85\columnwidth]{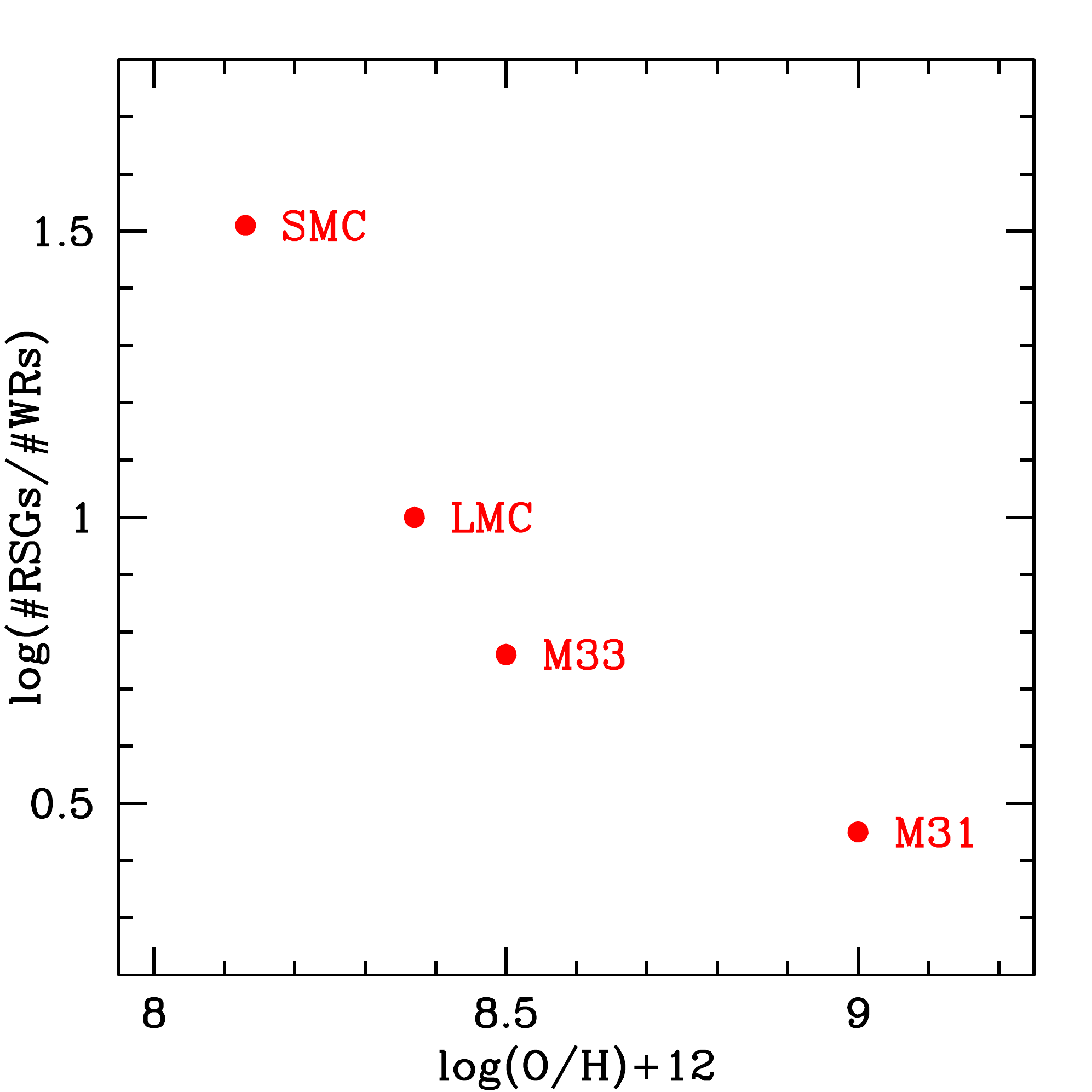}
\caption{\label{fig:rsg}We have counted the number of RSGs more luminous than $M_V\sim -5$ with 
$T_{\rm eff}<4000$ and compared that to the number of WRs as a function of metallicity.}
\end{center}
\end{figure}

An advantage of using WRs in an external galaxy for such studies is that the stars all lie at essentially the same distances; thus the second criterion (being volume-limited) is automatically met. Sadly, studies of the WR content of the Milky Way do not share this, as distances are highly uncertain, greatly magnifying the problems with other selection effects.  Nor is it clear what we learn by detecting a few WCs in more distant galaxies, so this review will be restricted to the WRs in Local Group.

\section{Past Surveys for WRs}

One of the testable predictions of massive star evolutionary models is the relative number of WC- and WN-type WRs as a function of metallicity.  In a completely mixed-age population (such as what we observe by averaging over many star-forming regions in a galaxy) we would naively expect this ratio to increase with metallicity according to the Conti Scenario, as at high metallicity we expect stars of somewhat lower masses will suffer sufficient mass-loss to become WC stars.  At lower metallicities the mass limit for evolving to the WC stage should be higher. And indeed, when the first author was a graduate student, this was known to be the case, with the WC to WN ratio changing from about 1:7 in the low metallicity SMC and 1:1 in the Milky Way, while the LMC, at intermediate metallicity, had an intermediate ratio (1:4.5) \citep[see e.g.][]{DanyPeter80}.  Shortly after that, \citet{MasseyConti83} confirmed that there was a galactrocentric gradient in the WC to WN ratio in M33, consistent, perhaps, with its metallicity gradient.  And, our good colleagues \citep{ms83, ms87} were busy discovering WRs in M31, almost all of WC type, again consistent (one might imagine) with M31's super-solar metallicity \citep{ZKH94}.

\begin{figure}[H]
\begin{center}
\includegraphics[width=0.9\columnwidth]{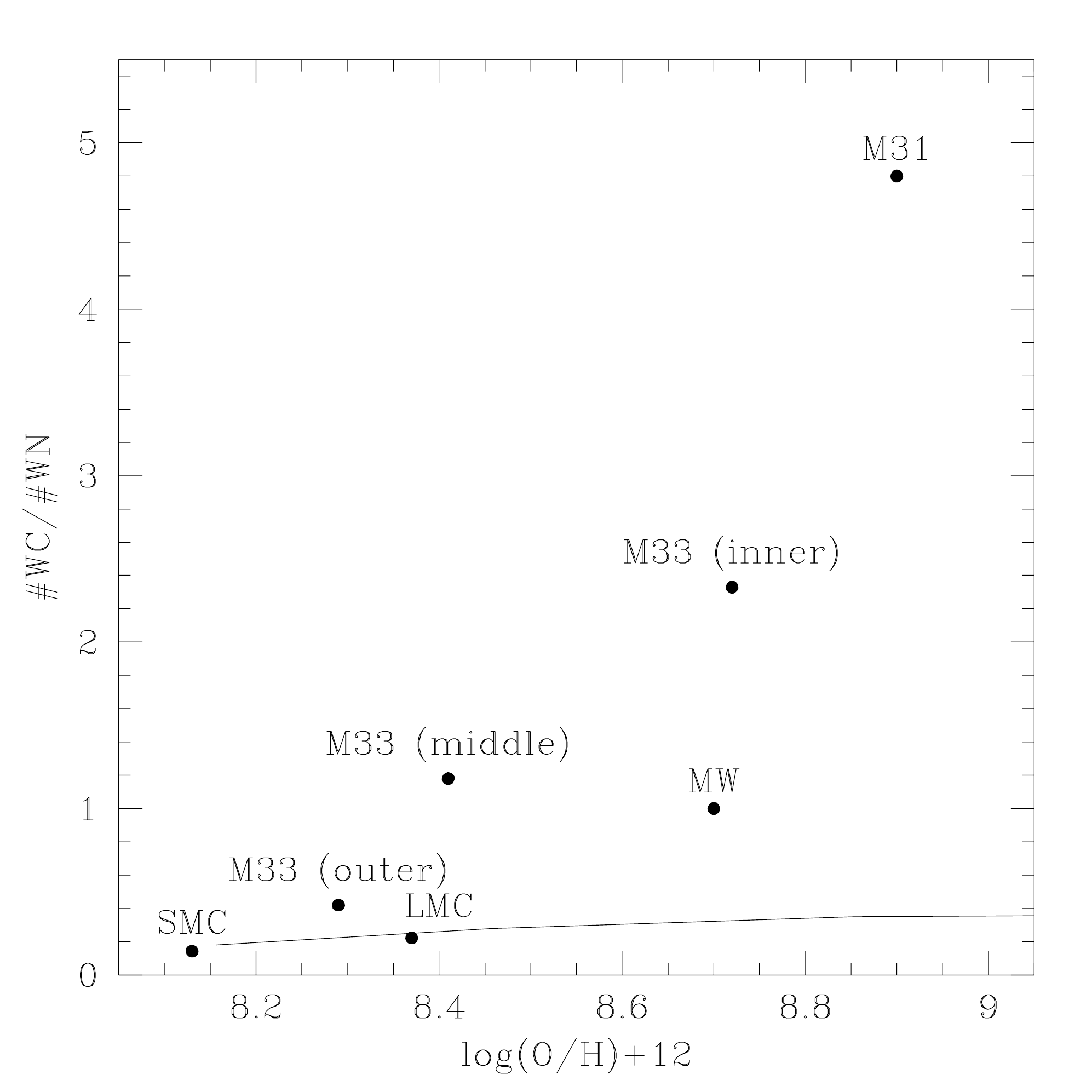}
\caption{\label{fig:wcwnold} The relative number of WC and WN stars as a function of metallicity as we knew from photographic studies, compared to the predictions of the Geneva evolutionary models from \citet{MeynetMaeder05}.}
\end{center}
\end{figure}

The problem came when one attempted to compare the actual numbers with the predictions of the stellar evolutionary models.  Recall that the predicted ratio isn't simply due to the lowering of the mass limits with increased metallicities: the relative lifetimes of the WC and the WN stage also play a critical role.  And when these are folded in with the mass limits, one finds a rather disturbing disagreement between the data (such as we knew it in the early 1980s) and the models, as Fig.~\ref{fig:wcwnold} shows dramatically.   (For convenience, we have adopted the predictions from the rotating models of Meynet \& Maeder 2005, although we had nothing so advanced at the time.)  We find that there is a factor of 13 difference between what the observations and models predict for the WC to WN ratio in M31!  
Clearly something was very wrong.  Interpreted naively, this would suggest a horrendous problem with the model mass limits and/or WR lifetimes.

However, there was another possibility.  The strongest optical emission line (C\,{\sc iii} $\lambda 4650$) in WC-type WRs is, on average, much stronger than that of the  strongest
emission line (He\,{\sc ii} $\lambda 4686$) in WN-type WRs \citep[see, e.g.,][]{MCA87,CM89}.  What, then, if the surveys of the galaxies beyond the Magellanic Clouds were just very biased towards WCs?  In all honesty, this thought did occur to us at the time; \citet{MasseyConti83} discuss this as a possibility, but conclude that the amount of incompleteness would have to be too great to explain the discrepancy even between the outer regions of M33 and the SMC/LMC.

\subsection{A Brief Journey Through Time}

To understand why the WC/WN ratios of the higher metallicities systems agreed so poorly with the models let us consider what we are calling the three eras of Local Group WR surveys.

\paragraph{\bf The First Era: Photographic.} The data in Fig.~\ref{fig:wcwnold} came photographic surveys.  These began with the Henry Draper Catalogue (1918-1924) and its extension (1925-1936), both general objective prism surveys which found many of the bright Galactic and LMC WRs.   These were followed by general objective prism surveys of the Magellanic Clouds, \citep{Sanduleak, AV75}, which found not only OB stars but also a wealth of WRs. There followed the first searches specifically for WRs, namely  \citet{AB79SMC,AB79LMC,AB80}.  They did a very clever thing: they took photographic objective prism images of fields in the Magellanic Clouds, but added a 120\AA\ wide filter that centered at 4650\AA\ in order to isolate only objects with C\,{\sc iii} $\lambda 4650$ and/or He\,{\sc ii} $\lambda 4686$ emission.  This had the advantage of cutting down on crowding, and also greatly reduced the sky contribution, allowing them to go much deeper than would otherwise have been possible.   The photographic era ended with WR surveys being extended to the more distant members of the Local Group, M33 and M31.
\citet{WrayCorso72} used an on-band He\,{\sc ii}/C\,{\sc iii} filter and a continuum filter to take images of two fields in M33, identifying the first WRs beyond the Magellanic Clouds.  The same technique was used by \citet{MasseyConti83} for a more extensive survey of M33, and by \citet{ms83,ms87} for two fields in M31.

\paragraph{\bf The Second Era: Optimal filters and Small CCD fields.}  With the advent of CCDs, it was no longer necessary to blink on-band and off-band photographic plates trying to spot the change in brightness that would demark the presence of a WR star.
Instead, quantitative photometry could be used to measure the brightness of all the objects on a CCD frame, and the magnitude differences compared to the photometric errors to find statistically significant candidates.  \citet{AM85} were the first to apply this technique, searching for WR stars in the nearby irregular galaxies NGC 6822 and IC 1613, as well as two test fields in M33.
For this work they designed a new filter system, consisting of 50\AA-wide bandpasses centered on C\,{\sc iii} $\lambda 4650$, He\,{\sc ii} $\lambda 4686$, and neighboring continuum at 4750\AA.  This system was designed to be optimized for the detection of WRs, using the extensive spectrophotometry of the normal (non-WR) stars from \citet{JHC} and WRs from \citet{Massey84} and \citet{MCLMC83}.  They discovered three more WNs in NGC~6822 \citep{MCA87} in addition to the one previously known \citep{Westerlund83}, but more importantly their study identified five new WN stars in a M33 field previously known only to contain six WCs and two WNs.  (Compare Table 5 in Armandroff \& Massey1985 to Table 2 in Massey et al.\ 1987.)  Thus, in this one field the WC/WN ratio changed from 3.0 to 0.9.  The significance of this was not immediately apparent, but further studies hammered home the point that the photographic studies had been woefully incomplete for WNs.   In particular, \citet{MAC86} surveyed eight small fields in M31 for WRs, identifying new WR candidates, many of which were subsequently spectroscopically confirmed as WRs \citep{MCA87,AM91}.  \citet{MJ98} extended these studies to include additional fields in M33, provided catalogs of all WRs beyond the Magellanic Clouds, and extensively discussed the selection biases against WNs.  Their conclusion was that the photographic studies had been 50\% incomplete for WNs, a number we might consider now to have been conservative.  These studies also demonstrated the need for spectroscopic confirmation of any photometrically-detected WR candidates: none of the new  IC~1613 WR candidates found by \citet{AM85} turned out to be real.  Similarly, \citet{Royer98} designed their own photometric system aimed at identifying WRs, and used this to announce  the discovery of the first WC9 stars detected in another galaxy \citep{Royer01}. When observed spectroscopically, none of the WC9 candidates proved to be WRs \citep{Chowder03}.

\paragraph{\bf The Third Era: Large CCDs and Image Subtraction Techniques.} Although CCDs were much more sensitive than photographic plates, and allowed quantitative assessment of the candidates, the early chips were {\it tiny} compared to plates, and the areal coverages about large enough to include a single OB association in one of these galaxies.  The introduction of larger chips and mosaic cameras provided the means to finally complete surveys for WRs in M33 \citep{NeugentM33} and even M31 \citep{NeugentM31}.  
Furthermore, our supernovae colleagues had spent years developing powerful image subtraction techniques.  Combined with photometry, these greatly reduced the number of false positives we had typically experienced by just using photometry.  (When one is considering 10,000 stars, a 3$\sigma$ criterion will lead to 15 spurious detections!) With these deeper surveys what has happened to the large discrepancy with the models shown in Fig.~\ref{fig:wcwnold}?  The improved data are shown in Fig.~\ref{fig:wcwnnew}.   

\begin{figure}[H]
\begin{center}
\includegraphics[width=0.9\columnwidth]{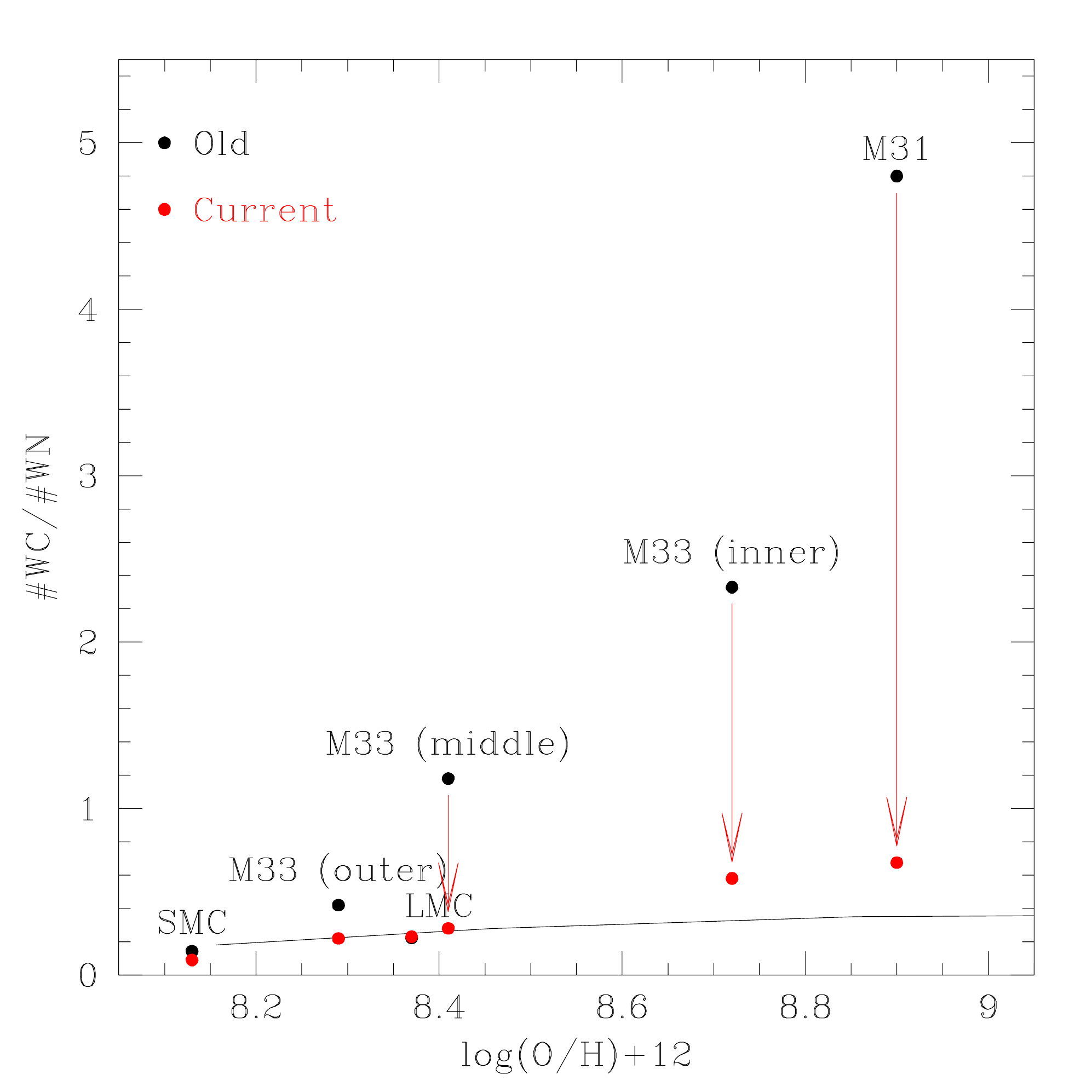}
\caption{\label{fig:wcwnnew}The relative number of WC and WN stars as a function of metallicity as we know today (red) compared to the older data shown in Fig.~\ref{fig:wcwnold}. The agreement with the models is significantly improved!}
\end{center}
\end{figure}

What haven't  we included in Fig.~\ref{fig:wcwnnew}?  First, we have ignored NGC 6822 and IC~1613 due to their very small number statistics: NGC 6822 has four WNs confirmed \citep[][and references therein]{MCA87}.   There is one WR star known in IC~1613; this is of WO-type. We have also not included any numbers for the Milky Way, as we suspect that these are still incomplete, and defining a volume-limited sample from the distances derived from spectral types to contain too many biases.

The galaxy most conspicuous by its absence is IC~10.  IC~10 is a metal-poor ($\log$ O/H+12 $\sim$8.3) irregular Local Group galaxy, chracterized as a starburst due to the surprising discovery of 15 WRs by \citet{MCA92} and \citet{MA95}.  This meant that IC~10 had $\sim5\times$ the surface density of WRs as the SMC, comparable to that found in the most active OB associations \citep{MasseyHolmes}.  Even more surprising was that the WC/WN ratio in IC~10 was very high, $\sim$2, despite the fact that the metallicity was low;  one would expect a value about 0.2.   Although it seemed unlikely that the answer was that there were a very large number of WNs unaccounted for, nine of the additional candidates found by \citet{Royer01} were spectroscopically confirmed by \citet{Chowder03}, demonstrating the incompleteness of the \citet{MCA92} survey.   Using deeper images, \citet{MasseyHolmes} in fact found a large number of additional candidates, suggesting that the total WR population of IC~10 might be as large as 100!  The WC/WN ratio to now down to 1.2, and many dozens of candidates still await spectroscopy. 

What else have we learned from the study of WRs in nearby galaxies?  For one thing, the relative number of early and late WNs  change with metallicity, as does the the relative number of early and late WCs (Table 1).  For the WCs we think we understand this purely as a metallicity effect on the spectral appearance \citep{Chowder02}, but for the WNs, something more complicated is going on.


\begin{table}[H]
\begin{center} 
\captionabove{Early and Late-Type WRs}
\label{table:earlylate}
\begin{tabular}{lccc}
\toprule
Galaxy & log O/H &  WNE/ & WCE/ \\ 
       & +12       & WNL &    WCL \\
\midrule 
SMC     & 8.1 & 4.5 & $\infty$ \\
M33out & 8.3 & 4.4 &  $\infty$ \\
LMC      & 8.4 &1.6 & 12 \\
M33mid & 8.4 & 1.2 &1.8 \\
M33in & 8.7    &  0.9: & 1.3 \\
M31 & 8.9 &  1.2 &  0.4 \\
\bottomrule
\end{tabular}
\end{center}
\end{table}

\section{A Modern Survey for WRs in the Magellanic Clouds}

During the 2013 Rhodes meeting the three authors decided that we had to conduct a modern search for WRs in the Magellanic Clouds.  Although the prevailing wisdom was that the WR content of the Clouds was mostly known, 7 new LMC WRs had been found accidentally since the \citet{BAT99} catalog.  All but one of these were WNs, suggesting that the WC/WN ratio might be biased even in the Magellanic Clouds.  The seventh star was found by ourselves, and was a very strong-lined WO-type star, only the second known in the LMC \citep{NeugentWO}.  So, it seemed as if we still had some work to do.  We designed a multi-year project, in which we would apply the same successful techniques used in M33 and M31 by \citet{NeugentM33} and \citet{NeugentM31}, with the imaging done on the Las Campanas Swope, and the followup spectroscopy with Magellan. 

We have now finished the second year of the survey; our results are described by \citet{MCWRI} and \citet{MCWRII}.  With 60\% of the survey complete, we have discovered 13 new WRs in the LMC, plus a variety of other previously unknown interesting emission-line objects.  However, the most interesting finding is that 8 of these 13 WRs are of a type never before recognized, stars that would naively be classified as WN3+O3~V.  These stars are, however, too faint (by several magnitudes) to harbor an O3~V star, and our modeling has shown that we can reproduce both the emission and absorption lines with a single set of physical parameters \citep{MCWRI}.   In this conference, \citet{NeugentWN3O3} discuss these stars in detail.   Here we would like to comment on a simple question: why haven't we found these stars elsewhere.  

Surveys such as ours are not just flux-limited.  As emphasized by \citet{MJ98}, what matters is the magnitude {\it difference} $\Delta$m between the on-band filter and the off-band filter.  Thus, even though an Of-type star has a great deal of flux in the He\,{\sc ii} $\lambda 4686$ emission line, they are relatively difficult to detect because they are also bright in the continuum.  In other words, the detection limit is sensitive to the equivalent width of the line as a function of continuum flux. 
In Fig.~\ref{fig:complete} (left) we see the WRs and Of-stars we successfully detected in our survey, along with our 3$\sigma$ and 5$\sigma$ detection limits.   

\begin{figure*}[t]
\begin{center}
\includegraphics[width=0.85\columnwidth]{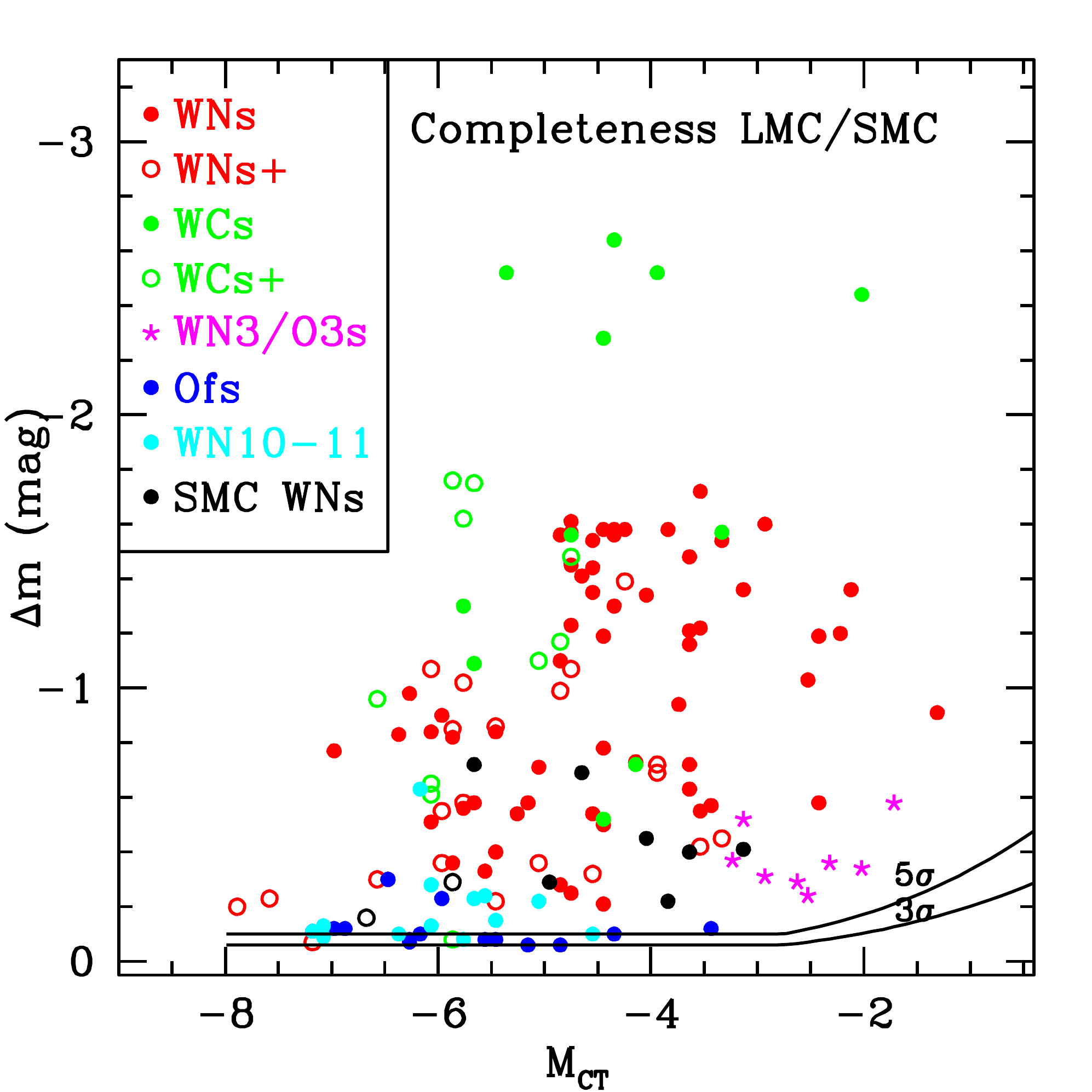}
\includegraphics[width=0.85\columnwidth]{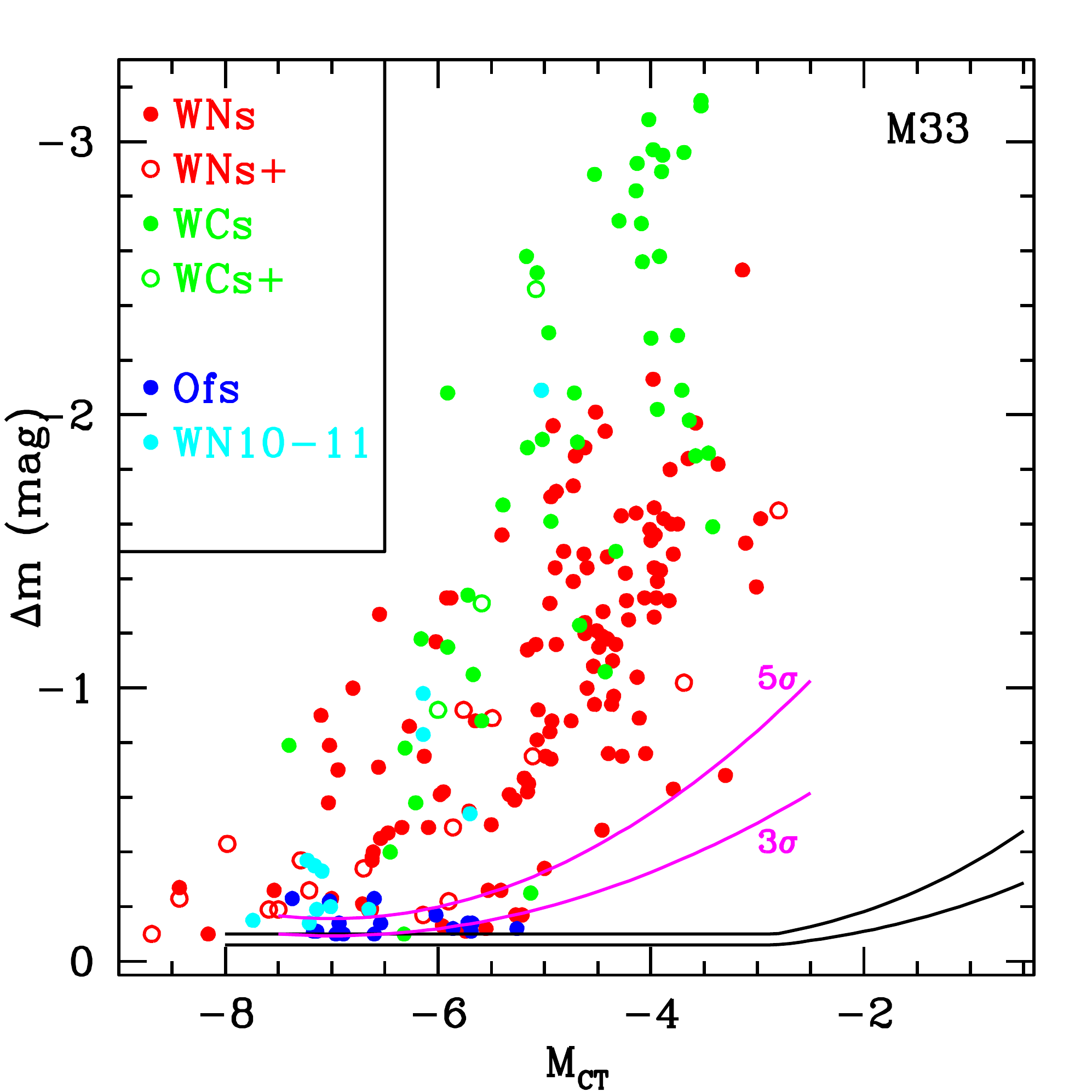}
\caption{\label{fig:complete} The magnitude difference $\Delta$ m is plotted against the absolute magnitude $M_{\rm CT}$ for WRs in the LMC/SMC (left) and in M33 (right).  The 5$\sigma$ and 3$\sigma$ detection limits are shown.} 
\end{center}
\end{figure*}

Fig.~\ref{fig:complete} (left) shows that our survey goes several magnitudes deeper than the faintest WN3/O3s we find, and so there aren't even fainter ones that we are missing.  Note too that one or two of the SMC WNs are in a similar region of the diagram.  All but one of the SMC WRs show absorption lines, and for many years there has been speculation that all of these stars are binaries, yet only four have orbit solutions. \citet{PotsSMC} has now shown that the absorption and emission can be modeled with a single set of physical parameters.  The SMC WNs are more luminous visually than our WN3/O3s, and thus are not the same thing, but likely are related. 

Why haven't WN3/O3s been found elsewhere in the Local Group?  They probably don't form at high metallicities as no Milky Way ones are known, so it's not surprising that we haven't found them in M31. But what about M33? Earlier we've said that surveys need to be ``sensitive enough to detect the weakest-lined WRs at the faintest magnitude limits one expects to find them."  But what if we don't know how faint that is? The LMC WN3/O3s are quite faint, with $M_V$$\sim$$-2.5$ to $-3.0$.  Do we go sufficiently deep in the M33 survey to find them?  As Fig.~\ref{fig:complete} (right) shows, the answer is ``no."  We should have found most of the other M33 WRs, but to find WN3/O3s will require a bit more work, and we are planning deeper imaging to find such stars.

We gratefully acknowledge support of this project by Lowell Observatory and the National Science Foundation through AST-1008020.

\bibliographystyle{aa} 
\bibliography{myarticle}

\begin{thebibliography}{41}
\expandafter\ifx\csname natexlab\endcsname\relax\def\natexlab#1{#1}\fi

\bibitem[{{Armandroff} \& {Massey}(1985)}]{AM85}
{Armandroff}, T.~E. \& {Massey}, P. 1985, \apj, 291, 685

\bibitem[{{Armandroff} \& {Massey}(1991)}]{AM91}
{Armandroff}, T.~E. \& {Massey}, P. 1991, \aj, 102, 927

\bibitem[{{Azzopardi} \& {Breysacher}(1979{\natexlab{a}})}]{AB79SMC}
{Azzopardi}, M. \& {Breysacher}, J. 1979{\natexlab{a}}, \aap, 75, 120

\bibitem[{{Azzopardi} \& {Breysacher}(1979{\natexlab{b}})}]{AB79LMC}
{Azzopardi}, M. \& {Breysacher}, J. 1979{\natexlab{b}}, \aap, 75, 243

\bibitem[{{Azzopardi} \& {Breysacher}(1980)}]{AB80}
{Azzopardi}, M. \& {Breysacher}, J. 1980, \aaps, 39, 19

\bibitem[{{Azzopardi} {et~al.}(1975){Azzopardi}, {Vigneau}, \&
  {Macquet}}]{AV75}
{Azzopardi}, M., {Vigneau}, J., \& {Macquet}, M. 1975, \aaps, 22, 285

\bibitem[{{Breysacher} {et~al.}(1999){Breysacher}, {Azzopardi}, \&
  {Testor}}]{BAT99}
{Breysacher}, J., {Azzopardi}, M., \& {Testor}, G. 1999, \aaps, 137, 117

\bibitem[{{Conti}(1975)}]{Conti75}
{Conti}, P.~S. 1975, Memoires of the Societe Royale des Sciences de Liege, 9,
  193

\bibitem[{{Conti} \& {Massey}(1989)}]{CM89}
{Conti}, P.~S. \& {Massey}, P. 1989, \apj, 337, 251

\bibitem[{{Crowther} {et~al.}(2002){Crowther}, {Dessart}, {Hillier}, {Abbott},
  \& {Fullerton}}]{Chowder02}
{Crowther}, P.~A., {Dessart}, L., {Hillier}, D.~J., {Abbott}, J.~B., \&
  {Fullerton}, A.~W. 2002, \aap, 392, 653

\bibitem[{{Crowther} {et~al.}(2003){Crowther}, {Drissen}, {Abbott}, {Royer}, \&
  {Smartt}}]{Chowder03}
{Crowther}, P.~A., {Drissen}, L., {Abbott}, J.~B., {Royer}, P., \& {Smartt},
  S.~J. 2003, \aap, 404, 483

\bibitem[{{Hainich} {et~al.}(2015){Hainich}, {Pasemann}, {Todt}, {Shenar},
  {Sander}, \& {Hamann}}]{PotsSMC}
{Hainich}, R., {Pasemann}, D., {Todt}, H., {et~al.} 2015, \aap, in press

\bibitem[{{Jacoby} {et~al.}(1984){Jacoby}, {Hunter}, \& {Christian}}]{JHC}
{Jacoby}, G.~H., {Hunter}, D.~A., \& {Christian}, C.~A. 1984, \apjs, 56, 257

\bibitem[{{Maeder} {et~al.}(1980){Maeder}, {Lequeux}, \&
  {Azzopardi}}]{Maeder80}
{Maeder}, A., {Lequeux}, J., \& {Azzopardi}, M. 1980, \aap, 90, L17

\bibitem[{{Massey}(1984)}]{Massey84}
{Massey}, P. 1984, \apj, 281, 789

\bibitem[{{Massey}(2003)}]{MasseyARAA}
{Massey}, P. 2003, \araa, 41, 15

\bibitem[{{Massey} \& {Armandroff}(1995)}]{MA95}
{Massey}, P. \& {Armandroff}, T.~E. 1995, \aj, 109, 2470

\bibitem[{{Massey} {et~al.}(1986){Massey}, {Armandroff}, \& {Conti}}]{MAC86}
{Massey}, P., {Armandroff}, T.~E., \& {Conti}, P.~S. 1986, \aj, 92, 1303

\bibitem[{{Massey} {et~al.}(1992){Massey}, {Armandroff}, \& {Conti}}]{MCA92}
{Massey}, P., {Armandroff}, T.~E., \& {Conti}, P.~S. 1992, \aj, 103, 1159

\bibitem[{{Massey} \& {Conti}(1983{\natexlab{a}})}]{MasseyConti83}
{Massey}, P. \& {Conti}, P.~S. 1983{\natexlab{a}}, \apj, 273, 576

\bibitem[{{Massey} \& {Conti}(1983{\natexlab{b}})}]{MCLMC83}
{Massey}, P. \& {Conti}, P.~S. 1983{\natexlab{b}}, \apj, 264, 126

\bibitem[{{Massey} {et~al.}(1987){Massey}, {Conti}, \& {Armandroff}}]{MCA87}
{Massey}, P., {Conti}, P.~S., \& {Armandroff}, T.~E. 1987, \aj, 94, 1538

\bibitem[{{Massey} \& {Holmes}(2002)}]{MasseyHolmes}
{Massey}, P. \& {Holmes}, S. 2002, \apjl, 580, L35

\bibitem[{{Massey} \& {Johnson}(1998)}]{MJ98}
{Massey}, P. \& {Johnson}, O. 1998, \apj, 505, 793

\bibitem[{{Massey} {et~al.}(2015){Massey}, {Neugent}, \& {Morrell}}]{MCWRII}
{Massey}, P., {Neugent}, K.~F., \& {Morrell}, N. 2015, \apj, 807, 81

\bibitem[{{Massey} {et~al.}(2014){Massey}, {Neugent}, {Morrell}, \&
  {Hillier}}]{MCWRI}
{Massey}, P., {Neugent}, K.~F., {Morrell}, N., \& {Hillier}, D.~J. 2014, \apj,
  788, 83

\bibitem[{{Meynet} \& {Maeder}(2005)}]{MeynetMaeder05}
{Meynet}, G. \& {Maeder}, A. 2005, \aap, 429, 581

\bibitem[{{Moffat} \& {Shara}(1983)}]{ms83}
{Moffat}, A.~F.~J. \& {Shara}, M.~M. 1983, \apj, 273, 544

\bibitem[{{Moffat} \& {Shara}(1987)}]{ms87}
{Moffat}, A.~F.~J. \& {Shara}, M.~M. 1987, \apj, 320, 266

\bibitem[{{Neugent} \& {Massey}(2011)}]{NeugentM33}
{Neugent}, K.~F. \& {Massey}, P. 2011, \apj, 733, 123

\bibitem[{{Neugent} {et~al.}(2012{\natexlab{a}}){Neugent}, {Massey}, \&
  {Georgy}}]{NeugentM31}
{Neugent}, K.~F., {Massey}, P., \& {Georgy}, C. 2012{\natexlab{a}}, \apj, 759,
  11

\bibitem[{{Neugent} {et~al.}(2015){Neugent}, {Massey}, {Hillier}, \&
  {Morrell}}]{NeugentWN3O3}
{Neugent}, K.~F., {Massey}, P., {Hillier}, D.~J., \& {Morrell}, N. 2015, in
  Wolf-Rayet Star Workshop, ed. W.~{Hamann}, A.~{Sander}, \& H.~{Todt}

\bibitem[{{Neugent} {et~al.}(2012{\natexlab{b}}){Neugent}, {Massey}, \&
  {Morrell}}]{NeugentWO}
{Neugent}, K.~F., {Massey}, P., \& {Morrell}, N. 2012{\natexlab{b}}, \aj, 144,
  162

\bibitem[{{Royer} {et~al.}(2001){Royer}, {Smartt}, {Manfroid}, \&
  {Vreux}}]{Royer01}
{Royer}, P., {Smartt}, S.~J., {Manfroid}, J., \& {Vreux}, J.-M. 2001, \aap,
  366, L1

\bibitem[{{Royer} {et~al.}(1998){Royer}, {Vreux}, \& {Manfroid}}]{Royer98}
{Royer}, P., {Vreux}, J.-M., \& {Manfroid}, J. 1998, \aaps, 130, 407

\bibitem[{{Sanduleak}(1970)}]{Sanduleak}
{Sanduleak}, N. 1970, Contributions from the Cerro Tololo Inter-American
  Observatory, 89

\bibitem[{{Smith} \& {Owocki}(2006)}]{SmithOwocki06}
{Smith}, N. \& {Owocki}, S.~P. 2006, \apjl, 645, L45

\bibitem[{{Vanbeveren} \& {Conti}(1980)}]{DanyPeter80}
{Vanbeveren}, D. \& {Conti}, P.~S. 1980, \aap, 88, 230

\bibitem[{{Westerlund} {et~al.}(1983){Westerlund}, {Azzopardi}, {Breysacher},
  \& {Lequeux}}]{Westerlund83}
{Westerlund}, B.~E., {Azzopardi}, M., {Breysacher}, J., \& {Lequeux}, J. 1983,
  \aap, 123, 159

\bibitem[{{Wray} \& {Corso}(1972)}]{WrayCorso72}
{Wray}, J.~D. \& {Corso}, G.~J. 1972, \apj, 172, 577

\bibitem[{{Zaritsky} {et~al.}(1994){Zaritsky}, {Kennicutt}, \&
  {Huchra}}]{ZKH94}
{Zaritsky}, D., {Kennicutt}, Jr., R.~C., \& {Huchra}, J.~P. 1994, \apj, 420, 87

\end{thebibliography}

\end{multicols}

\end{contribution}


\end{document}